\documentclass[usenatbib]{mn2e}
\usepackage{natbibmnfix,graphicx,times}

\newcommand{\kel}{\mbox{ K}}

\newcommand{\Mpc}{\mbox{ Mpc}}
\newcommand{\Mpcden}{\mbox{ Mpc}^{-3}}

\newcommand{\msun}{\mbox{ M$_\odot$}}

\newcommand{\hunits}{\mbox{ km s$^{-1}$ Mpc$^{-1}$}}
\newcommand{\kms}{\mbox{ km s$^{-1}$}}
\newcommand{\bq}{\begin{equation}}
\newcommand{\eq}{\end{equation}}
\newcommand{\bqa}{\begin{eqnarray}}
\newcommand{\eqa}{\end{eqnarray}}

\newcommand{\bxhi}{\bar{x}_{\rm HI}}

\newcommand{\bxio}{\bar{x}_i}
\newcommand{\lya}{Ly$\alpha$ }
\newcommand{\lyans}{Ly$\alpha$}
\newcommand{\hii}{HII }
\newcommand{\mmin}{m_{\rm min}}
\newcommand{\fcoll}{f_{\rm coll}}
\newcommand{\deriv}{{\rm d}}

\newcommand{\apj}{ApJ}
\newcommand{\apjl}{ApJ}
\newcommand{\apjs}{ApJS}

\newcommand{\aj}{AJ}
\newcommand{\mnras}{MNRAS}
\newcommand{\pasj}{PASJ}
\newcommand{\physrep}{Physics Reports}
\newcommand{\nat}{Nature}

\title[Reionization and \lya galaxy surveys]{The effects of reionization on \lya galaxy surveys}

\author[Furlanetto et al.]{Steven R. Furlanetto,$^1$\thanks{Email: sfurlane@tapir.caltech.edu} Matias Zaldarriaga,$^{2,3}$ \& Lars Hernquist$^2$ \\
$^1$Division of Physics, Mathematics, \& Astronomy; California Institute of Technology; Mail Code 130-33; Pasadena, CA 91125 \\
$^2$Harvard-Smithsonian Center for Astrophysics, 60 Garden St., Cambridge, MA 02138 \\
$^3$Jefferson Laboratory of Physics; Harvard University; Cambridge, MA 02138}

\begin{document}

\maketitle

\begin{abstract}
Searches for \lya emission lines are among the most
effective ways to identify high-redshift galaxies.  They are
particularly interesting because they probe not only the galaxies
themselves but also the ionization state of the intergalactic medium
(IGM): nearby neutral gas efficiently absorbs \lya photons.  The
observed line strengths depend on the amount by which each photon is
able to redshift away from line center before encountering neutral gas
and hence on the size distribution of \hii regions surrounding the
sources.  Here, we use an analytic model of that size distribution to
study the effects of reionization on the luminosity function of
\lyans-emitters and their observed spatial distribution.  Our model
includes the clustering of high-redshift galaxies and thus contains
ionized bubbles much larger than those expected around isolated
galaxies.  As a result, \lyans-emitting galaxies remain visible earlier
in reionization: we expect the number counts to decline by only a
factor $\sim 2$ (or 10) when the mean ionized fraction falls to $\bxio \sim
0.75$ (or 0.5) in the simplest model.  Moreover, the absorption is not uniform across the sky:
galaxies remain visible only if they sit inside large bubbles, which
become increasingly rare as $\bxio$ decreases.  Thus, the size
distribution also affects the apparent clustering of \lyans-selected
galaxies.  On large scales, it traces that of the large bubbles, which
in our model are more biased than the galaxies.  On small scales, the
clustering increases rapidly as $\bxio$ decreases because large \hii
regions surround strong galaxy overdensities, so a survey
automatically selects only those galaxies with neighbours.  The
transition between these two regimes occurs at the characteristic
bubble size.  Hence, large \lya galaxy surveys have the potential to
measure directly the size distribution of \hii regions during
reionization.

\end{abstract}


\section{Introduction}
\label{intro}

The reionization of hydrogen in the intergalactic medium (IGM) is a
landmark event, because it defines the moment at which structure
formation affected every baryon in the Universe.  In the past few
years, astronomers have made enormous strides in understanding this
transition.  The rapid onset of \citet{gunn65} absorption in the
spectra of $z>6$ quasars \citep{becker01,fan02,white03} indicates that
reionization probably ended at $z \sim 6$ (albeit with large variance
among different lines of sight: \citealt{wyithe04-var,oh05}).  On the other hand, the
detection of a large optical depth to electron scattering for cosmic
microwave background photons \citep{kogut03} indicates (albeit with
relatively low confidence) that the process began at $z \ga 15$.
Reionization thus appears to be a complex process, a conclusion
strengthened by several other studies
\citep{theuns02-reion,wyithe04-prox,mesinger04}, so new methods to
study the ``twilight zone" of reionization are crucial.

One particularly intriguing technique is to search for \lyans-emitting
galaxies in the high-redshift universe (e.g.,
\citealt{hu02-lya,rhoads04,santos04-obs,stanway04, taniguchi05}).
These surveys offer a number of advantages.  First, narrowband
searches reduce the sky background, especially if placed between the
bright sky lines that (nearly) blanket the near-infrared sky
(e.g., Barton et al. 2004).  Second, they efficiently select galaxies
at a known redshift (although with some contamination by
lower-redshift interlopers).  Third, they increase the signal-to-noise
by focusing on an emission line.  Beyond these practical advantages, such
surveys also constrain both the sources responsible for reionization
(and in particular young massive stars; \citealt{partridge67})
\emph{and} the ionization state of the IGM itself, because \lya
photons will be absorbed if they pass through neutral gas near the
galaxy.  This is a consequence of the enormous \lya optical depth of a
neutral IGM: $\tau_{\rm GP} \sim 6.5 \times 10^5 \, \bxhi \,
[(1+z)/10]^{3/2}$ \citep{gunn65}, so even those photons passing through
the damping wing of the \lya resonance will be absorbed
\citep{miralda98}.

Thus, as the IGM becomes more neutral, the \lya selection technique
will detect fewer and fewer objects (even after accounting for
cosmological evolution in their intrinsic abundance); the number of
such galaxies therefore measures the globally averaged ionized
fraction $\bxio$ \citep{madau00,haiman02-lya,santos04}.  Recently,
\citet{malhotra04} used these arguments to constrain the neutral
fraction at $z=6.5$ (see also \citealt{stern05}).  They compared
luminosity functions of \lyans-emitters at $z=5.7$ and $z=6.5$
(bracketing the time at which quasar spectra indicate that
reionization ends) and found no evolution in the number density over
this range.  By comparing to existing models of the \lya damping wing
absorption around isolated galaxies \citep{santos04}, they concluded
that this required $\bxio \ga 0.7$ at $z=6.5$.  \citet{haiman05-lya}
reached a similar conclusion using different models (but the same
luminosity function data).

The optical depth encountered by a galaxy's \lya photons depends
primarily on the extent of the \hii region surrounding it: the photons
redshift as they stream through the ionized gas (suffering little
absorption), so they are somewhere in the wings of the line by the
time they encounter the neutral gas.  Thus, the amount of absorption
depends sensitively on the size distribution of ionized bubbles during
reionization.  Existing analyses treat each galaxy in isolation
\citep{santos04,haiman05-lya}, so the \hii regions are rather small
even late in reionization.  However, both analytic models
\citep{barkana04-fluc} and simulations
\citep{ciardi03-sim,sokasian03,sokasian04} show that galaxies are
highly clustered at these high redshifts.  \citet[hereafter
FZH04]{furl04-bub} developed an analytic model to calculate the size distribution of ionized bubbles around overdensities of sources.  They showed that galaxy \hii regions overlap quickly,
growing much larger than expected in isolation (reaching
scales of $\ga 10$ comoving Mpc when $\bxio \ga 0.65$).  We describe
this model in \S \ref{fzh}.  Such large bubbles obviously reduce the
damping wing absorption.  \citet[hereafter FHZ04]{furl04-lya} showed
that clustering therefore allows \lya lines to be visible so long as
the galaxy sits inside a relatively large bubble (see also
\citealt{wyithe05-clus}).  In \S \ref{lumfcn} we show that the large
\hii regions expected during reionization reduce the \lya absorption
for a given $\bxio$, and we compute the expected decline in the number
counts of \lyans-selected galaxies throughout reionization.

Although this weakens the constraint on $\bxio$ given by
\citet{malhotra04}, it implies that the \lya selection technique
should remain viable deeper into reionization, making these surveys
even more exciting.  Moreover, the \emph{distribution} of bubble sizes
implies that the damping wing absorption is spatially variable: the
bubbles essentially modulate the selection function of the survey.
Thus, when large bubbles are rare (early in reionization), we expect
to find isolated clumps of \lyans-emitting galaxies separated by
apparent voids.  We show in \S \ref{bias} that the bubbles affect the
observed clustering pattern on both large and small scales:
\lyans-selected galaxies will show a scale-dependent bias, with the
magnitude depending on $\bxio$ and a break appearing at the
characteristic scale of the bubbles.  We discuss the implications of
these results in \S \ref{disc}.

In our numerical calculations, we assume a cosmology with
$\Omega_m=0.3$, $\Omega_\Lambda=0.7$, $\Omega_b=0.046$, $H=100 h
\hunits$ (with $h=0.7$), $n=1$, and $\sigma_8=0.9$, consistent with
the most recent measurements \citep{spergel03}.  Unless otherwise
specified, we quote distances in comoving units.

\section{The FZH04 Model}
\label{fzh}

\begin{figure}
\begin{center}
\resizebox{8cm}{!}{\includegraphics{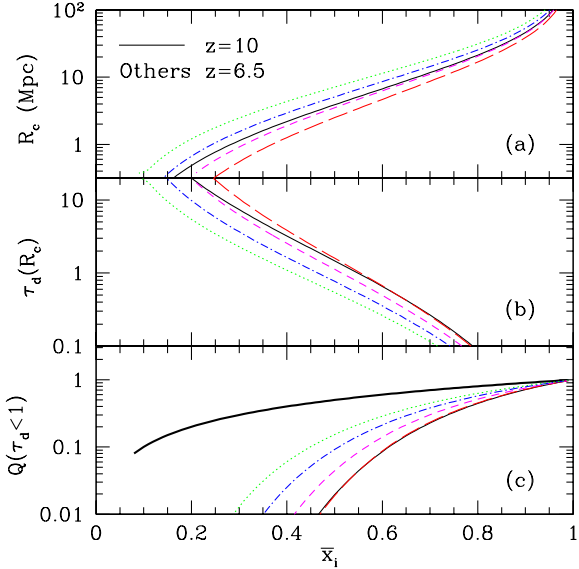}}\\%
\end{center}
\caption{\emph{(a)}: The characteristic bubble size $R_c$ throughout
reionization.  \emph{(b)}: The damping wing optical depth $\tau_d$ at
the center of a bubble of the size $R_c$.  \emph{(c)}: The fraction of
the universe filled by bubbles with $\tau_d<1$ (as measured from the
center of each bubble).  In each panel, the solid and long-dashed
curves take $\mmin=m_4$ at $z=10$ and $z=6.5$, respectively.  The
short-dashed, dot-dashed, and dotted curves take $z=6.5$ and
$\mmin/m_4=3,\,10$, and $30$, respectively.  The thick solid line in
the bottom panel shows $Q=\bxio$.}
\label{fig:td}
\end{figure}

Recent numerical simulations (e.g.,
\citealt{ciardi03-sim,sokasian03,sokasian04}) show that reionization
proceeds ``inside-out'' from high density clusters of sources to
voids.  We therefore associate \hii regions with large-scale
overdensities.  
We assume that a galaxy of total mass $m_{\rm gal}$ can ionize baryons
associated with a total mass of $\zeta m_{\rm gal}$, where $\zeta$ is
a constant that depends on (among other things) the efficiency of
ionizing photon production, the escape fraction of these photons from
the host galaxy, the star formation efficiency, and the mean number of
recombinations.  Each of these quantities is highly uncertain (e.g.,
massive Population III stars can dramatically increase the ionizing
efficiency; \citealt{bromm01-vms}), but at least to a rough
approximation they can be collapsed into this single efficiency
factor.  The criterion for a region to be ionized by the galaxies
contained inside it is then $\fcoll > \zeta^{-1}$, where $f_{\rm
coll}$ is the fraction of mass bound in haloes above some $\mmin$.
Unless otherwise specified, we will assume that $\mmin=m_4$,
corresponding to a virial temperature $T_{\rm vir}=10^4 \kel$ where
hydrogen line cooling becomes efficient.  In the extended
Press-Schechter model \citep{lacey93}, this places a condition on the
mean overdensity within a region of mass $m$, $\delta >
\delta_x(m,z)$.  FZH04 showed how to construct the mass function of
\hii regions from $\delta_x$ in an analogous way to the halo mass
function \citep{press74,bond91}.  We first approximate the threshold
$\delta_x$ as $\delta_x \approx B(m,z) \equiv B_0 + B_1 \sigma^2(m)$,
where $\sigma^2(m)$ is the variance of density fluctuations on the
scale $m$.  This linear approximation turns out to be quite accurate.
Then we can write the comoving number density of \hii regions with
masses in the range $m \pm \deriv m/2$ as \citep{sheth98,mcquinn05}:
\bqa n_b(m) \, \deriv m & = & \sqrt{\frac{2}{\pi}} \
\frac{\bar{\rho}}{m^2} \ \left| \frac{\deriv \ln \sigma}{\deriv \ln m}
\right| \ \frac{B_0(z)}{\sigma(m)} \nonumber \\ & & \times \exp \left[
- \frac{B^2(m,z)}{2 \sigma^2(m)} \right] \, \deriv m,
\label{eq:nbm}
\eqa
where $\bar\rho$ is the mean density of the universe.  FZH04 showed
some examples of how the bubble sizes evolve throughout the early and
middle stages of reionization.  Several key properties of this model
deserve emphasis.  First, the bubbles are large: tens of comoving Mpc
for $\bxio \ga 0.65$.  This is because high-redshift galaxies cluster
strongly, so small mass overdensities translate to much larger
overdensities of galaxies.  There are other independent lines of
evidence suggesting large \hii regions during reionization
\citep{wyithe04-var,cen05}.  Furthermore, the bubbles attain a
well-defined characteristic size $R_c$ at any point during
reionization.  We show how $R_c$ evolves in
Figure~\ref{fig:td}\emph{a}.  The solid line is for $z=10$ and the
long-dashed line is for $z=6.5$ with $\mmin=m_4$.  The other curves
increase $\mmin$ at $z=6.5$.  Including only the massive galaxies
increases the bias of the underlying galaxy population and hence makes
the bubbles larger \citep{furl05-charsize}.  Finally, $n_b(m)$ is
fairly robust to the underlying parameters of galaxy formation.  For
example, Figure~\ref{fig:td}\emph{a} shows that $R_c$ varies only
weakly with $z$ (or equivalently $\zeta$) at a fixed $\bxio$.

By constructing $n_b(m)$ and the halo mass function $n_h(m)$ with the
same formalism, the FZH04 model allows a straightforward connection
between the two populations.  For example, according to the extended
Press-Schechter model the halo number density inside a bubble of mass
$m_b$ is (\citealt{lacey93}; FHZ04)
\bqa
n_h(m_h|m_b) & = & \sqrt{\frac{2}{\pi}} \, \frac{\bar{\rho}}{m_h^2} \, \left|
  \frac{\deriv \ln \sigma}{\deriv \ln m_h} \right| \frac{\sigma^2(m_h)[\delta_c(z)-B(m_b)]}{[\sigma^2(m_h)-\sigma^2(m_b)]^{3/2}} \nonumber \\
  & & \times  
  \exp \left\{-\frac{[\delta_c(z)-B(m_b)]^2}{2 [\sigma^2(m_h)-\sigma^2(m_b)]} \right\},
\label{eq:nhcond}
\eqa
where $\delta_c$ is the critical density for spherical collapse.

The FZH04 model includes recombinations in only a crude sense (as a
global average number of recombinations per baryon).
\citet{furl05-rec} showed how to incorporate them in a more physically
plausible manner.  For a bubble to grow, the (local) mean free path of an
ionizing photon must exceed its radius.  The mean free path is
determined by the spatial extent and number density of dense, neutral
blobs (where the recombination time is short).  Thus, as a bubble
grows, its internal radiation background must also increase to ionize each of
these blobs more deeply.  As a consequence, the recombination rate
inside the bubble also increases; \hii regions saturate when this
recombination rate exceeds the ionizing emissivity (see also
\citealt{miralda00}).  \citet{furl05-rec} showed that, for reasonable
models of the IGM density field, recombinations halt the evolution at
a well-defined radius $R_{\rm max}$; any bubbles nominally larger than
this size fragment into regions with $R \approx R_{\rm max}$.  (Note,
however, that this saturation radius describes the mean free path of
ionizing photons, not the extent of contiguous ionized gas; obviously
as $\bxio \rightarrow 1$ ionized gas does fill the universe.)
Unfortunately, calculating $R_{\rm max}$ requires knowledge of the IGM
density field during reionization, which we currently lack.  We will
therefore leave it as a free parameter, noting that $R_{\rm max} \ga
20 \Mpc$ if the density field is similar to that at $z \sim 2$--$4$
\citep{miralda00}, but minihaloes could in principle reduce it to
$R_{\rm max} \la 5 \Mpc$.

\section{The Evolving Luminosity Function}
\label{lumfcn}

We will now examine how the FZH04 $n_b(m)$ affects the visibility of
\lyans-selected galaxies throughout reionization.  A full model requires
two elements: a prescription for associating \lya line luminosity with
halo mass and a prescription for the damping wing absorption $\tau_d$.
The latter is straightforward: to a first approximation, it depends
only on the distance from the galaxy to the edge of its \hii region,
which determines the amount by which each photon redshifts away from
line center before encountering the neutral gas, and the mean ionized
fraction of the absorbing gas.  
The velocity dispersions of high-redshift galaxies are $\approx 12
(m/m_4)^{1/3} \kms$, while the velocity width of a bubble is $\sim 100
(R/{\rm Mpc}) \kms$.  Thus the damping wing absorption is nearly
constant across the line throughout most of reionization (for a
specific example, see Fig. 5 of \citealt{furl04-lya}).  We use the
formula of \citet{miralda98} assuming that gas with $\bxio$ extends
along a path length $\Delta z=0.5$ from the edge of the \hii region.
This choice is somewhat arbitrary, but our results are not sensitive
to it: $\tau_d$ changes by $\la 5\%$ so long as $\Delta z > 0.25$ and
only $\sim 20\%$ for $\Delta z=0.1$.  Thus, unless the mean ionized
fraction evolves extremely rapidly, it should be a reasonable
estimate.  On the other hand, because the damping wing integrates over
a large path length (effectively $\ga 50$ comoving Mpc) it is
relatively insensitive to fluctuations in the ionized fraction outside
of the host bubble.  For simplicity, we set $\tau_d$ equal to its
value at the center of each bubble.

Note that we ignore a number of potential complications here.  The
most obvious is resonant absorption by residual neutral gas within
each bubble.  FHZ04 (among many others) have included this effect; for
typical galaxies, the resonant absorption has $\tau_\alpha \ga 1$
blueward of the line center, which eliminates the flux on the blue
side of the line.  However, with this optical depth the absorption
from each volume element is quite narrow in redshift space (i.e. the
associated damping wings are negligible), so the effective optical
depth for photons that begin on the red side of the line is much
smaller than unity (see Fig. 5 of FHZ04 for an illustrative example).
Thus, to first order, resonant absorption eliminates about half of the
flux from each galaxy; we can incorporate this into our model by
simply reducing the assumed intrinsic luminosity of each galaxy by a
factor of two.  This approximation ignores the fact that the ionizing
radiation field is stronger around brighter galaxies, reducing the
effective resonant optical depth; this would imprint a trend with
galaxy mass that our model does not include.  However, a number of
other effects could mitigate this trend.  Most importantly, the
properties of individual galaxies, such as infall regions and winds,
move the relative velocities of the line and the absorbing gas
\citep{santos04}.  If winds, for example, move the emission to the red
side of the line -- as occurs in lower-redshift samples (e.g.,
\citealt{shapley03}) -- resonant absorption becomes completely
unimportant.  Moreover, with our large bubbles the average ionizing
background can make a non-negligible contribution.  The galaxy's
neighbours might also matter, because galaxies are clustered even
inside the bubles.  All of these effects would tend to decrease the
difference in resonant absorption between bright and faint galaxies;
thus we have chosen to ignore this aspect in our model.  We also
neglect the spatial distribution of galaxies within each bubble and
the clustering of bubbles (see \S \ref{bias}).

Computing the \lya line luminosity of each galaxy is much more
problematic.  
There is no simple analytic model for this quantity, although
\citet{ledelliou05a} have shown that semi-analytic models can match
the observed luminosity functions over a range of redshifts by
assuming a flat, top-heavy initial mass function and a constant escape
fraction of ionizing photons (see also \citealt{ledelliou05b}).  There
are two obvious hurdles: the \lya luminosity should vary in time with
the star formation rate and vary spatially with the distribution of
dust and other absorbing gas within the galaxy (particularly winds).
Simulations include the former effect (at least in principle), but at
lower redshifts the number density of star-forming galaxies in
simulations (which already include the time variability) exceeds the
number density of observed \lyans-emitters by about an order of
magnitude \citep{furl05-lyaem} even though the same simulations match
the history of cosmic star formation (\citealt{SH03,HS03}, but see
Nagamine et al. 2004b) and the observed Lyman-break population
\citep{nagamine04a,night05} reasonably well.  The remaining
discrepancy can reasonably be attributed to geometry; for example,
\citet{shapley03} find that only $\sim 20\%$ of Lyman-break galaxies
at $z=3$ have \lya emission lines, with a hint that lower-luminosity
galaxies tend to have stronger lines.  How these factors evolve with
redshift is not well-constrained.  
\citet{shapley03} find that younger galaxies are less likely to have
\lya emission lines.  On the other hand, direct comparison of the
number counts of continuum-selected and \lya line-selected galaxies
suggests that the fraction of galaxies with strong \lya lines may
increase with redshift. In the absence of better constraints, we will
assume that the \lya luminosity of each galaxy is $L_\alpha \propto
\zeta m_h \propto m_h$ so that -- if there were no IGM absorption -- a
particular survey could detect every galaxy with $m_h > m_{\rm
obs,min}$.  While clearly simplistic, this serves to illustrate our
point.  So long as the geometric effects eliminate a random fraction
of the population, they will not affect our conclusions.  More
worrying is the assumption that $\zeta=$constant.  We find, however,
that this does not qualitatively affect the argument, although it does
modify the resulting numerical constraints (see \S \ref{disc}).

Figure~\ref{fig:td} illustrates the major effects we expect from the
evolving bubble population.  The characteristic bubble size grows
rapidly with $\bxio$, decreasing the mean absorption suffered by a
typical galaxy (see Figure~\ref{fig:td}\emph{b}).  Thus the \lya
galaxy abundance should begin to fall significantly once $\bxio \sim
0.5$, beyond which $\tau_d(R_c) \ga 1$. However, even if $R_c$ is
small, some larger bubbles still exist.  Figure~\ref{fig:td}\emph{c}
shows the fraction of space $Q$ filled by bubbles with $\tau_d < 1$.
These large bubbles are crucial for observing sources at early times.
On the other hand, $Q \rightarrow \bxio$ near the end of reionization
because $R_c$ moves well beyond the required scale.

More precisely, equation (\ref{eq:nhcond}) allows us to compute the
distribution of galaxies within each bubble and hence the distribution
of $\tau_d$ (FHZ04 give a closely related calculation).  With the
assumption that $L_\alpha \propto m_h$, a galaxy of mass $m_h$ will
lie in the survey if
\bq
\tau_d < \ln \left( \frac{m_h}{m_{\rm obs,min}} \right).
\label{eq:tdampcond}
\eq   
Given a bubble of mass $m_b$, equation (\ref{eq:tdampcond}) implicitly
yields the minimum mass $m_{h,{\rm min}}$ for a member galaxy to lie
in our survey.  Thus the total number density of observable galaxies
is
\bq
n(>L) = \int \deriv m_b \, n_b(m_b) V_b \, \int_{m_{h,{\rm min}}}^{m_b/\zeta} \deriv m_h \, n_h(m_h|m_b),
\label{eq:lumfcn}
\eq
where $V_b$ is the bubble volume.

\begin{figure}
\begin{center}
\resizebox{8cm}{!}{\includegraphics{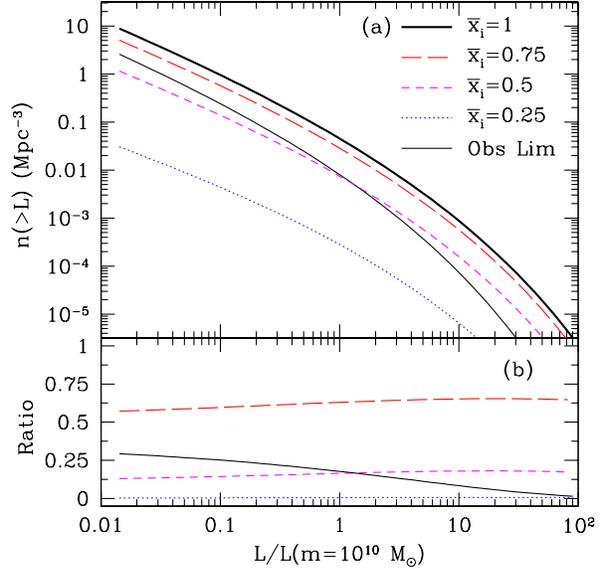}}\\%
\end{center}
\caption{\emph{(a)}: \lya luminosity functions at $z=6.5$.  The thick solid line shows the assumed
intrinsic function (with $L_\alpha \propto m_h$).  The thin solid line
shows the observational limit of \citet{malhotra04}, corresponding to
attenuation by a factor of three in comparison to the $z=5.7$
luminosity function.  The other curves show how $n(>L)$ is affected by
the bubble size distribution.  \emph{(b)}: Ratio of the attenuated to
intrinsic luminosity functions.}
\label{fig:z6lf}
\end{figure}

We show some examples of the resulting luminosity functions at $z=6.5$
in Figure~\ref{fig:z6lf}.  The thick solid curve in panel \emph{(a)}
is the assumed intrinsic luminosity function (neglecting absorption,
appropriate if $\bxio=1$).  The long-dashed, short-dashed, and dotted
curves assume $\bxio=0.75,\,0.5$, and $0.25$, respectively.  Panel
\emph{(b)} shows the ratio of $n(>L)$ after attenuation to its value
in a universe with $\bxio=1$.

We find that $n(>L)$ remains substantial even in the middle stages of
reionization: galaxies do not begin to disappear in large numbers
until $\bxio \la 0.5$.  We find the relative magnitude of the decline
to be nearly independent of redshift, because $n_b(m)$ at a given
$\bxio$ is also largely independent of redshift (and the mean optical
depth compensates for most of the remaining evolution; see
Fig.~\ref{fig:td}\emph{b}).  The number counts decline by $\sim 2$
when $\bxio=0.75$ and $\sim 10$ when $\bxio=0.5$ so long as $L_\alpha
\propto m_h$.  Interestingly, we find that -- at least in this simple
model -- the ratio of the attenuated and intrinsic luminosity
functions is nearly independent of galaxy luminosity.  This is a
result of two competing effects.  First, larger galaxies tend to
reside in larger bubbles (although this trend is much weaker than if
galaxies were isolated), suffering correspondingly weaker absorption.
However, the mass function steepens toward higher masses: thus if all
galaxies suffered identical absorption (as in the thin solid curves;
see below), the ratio would decrease sharply at large $L$ because a
fixed luminosity interval causes a much larger decrease in the number
counts.  Evidently these two effects nearly compensate.  If this
remains true in more sophisticated models, it would be useful in that
it removes the degeneracy of \lya galaxy tests of reionization with
the intrinsic luminosity limit of a survey.  
Note that, if resonant absorption is stronger around faint galaxies,
it would flatten the luminosity function and make bright galaxies
somewhat easier to see than faint galaxies (see the second paragraph
of this section).

\citet{malhotra04} compared the $z=5.7$ and $z=6.5$ luminosity
functions of \lyans-selected galaxies and found no evidence for
evolution; formally they ruled out a model in which each galaxy's line
is attenuated by a factor of three.  We show their constraint
(translated to our assumed luminosity function) by the thin solid
line.  To place a quantitative constraint, we must associate their
luminosity threshold with a halo mass.  The number density of detected
$z=6.5$ galaxies in their sample is $n \sim 10^{-4} \Mpcden$, which
would correspond to $m \sim 10^{11.5} \msun$ if there were a
one-to-one correspondence between haloes and \lyans-emitters.  Such a
large mass implies a weak constraint, $\bxio \ga 0.35$.  However,
comparison to simulations at lower redshifts indicates that fewer than
$\sim 10 \%$ of haloes actually host \lya lines \citep{furl05-lyaem}.
This would place the limit at $\bxio \ga 0.5$.  A more precise
constraint must await a more satisfactory model of \lyans-emitters,
including their detailed intrinsic properties (such as winds, which
can move the \lya line in velocity space and allow more transmission
than otherwise expected).  However, we note that our model should be
relatively robust to these details, because the length scales of our
bubbles are many times larger than those considered by, e.g.,
\citet{santos04}.

\begin{figure}
\begin{center}
\resizebox{8cm}{!}{\includegraphics{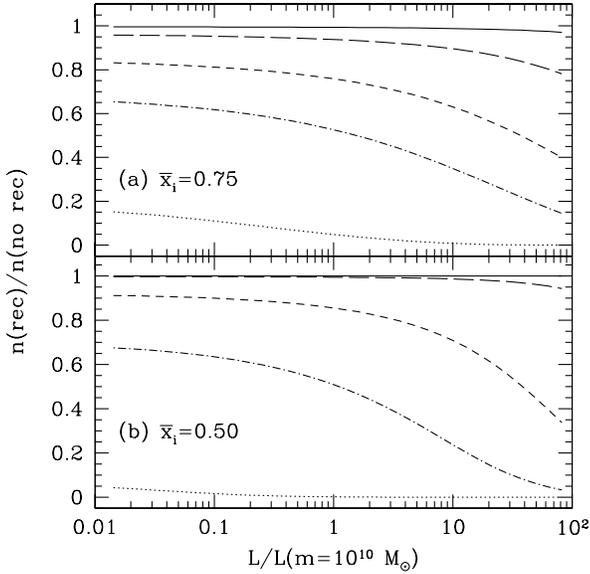}}\\%
\end{center}
\caption{Number density of galaxies above a luminosity $L$ at $z=6.5$ if recombinations impose a maximum bubble size $R_{\rm max}$, relative to their abundance without an imposed maximum size.  In each panel, the solid, long-dashed, short-dashed, dot-dashed, and dotted curves take $R_{\rm max}=20,\,10,\,5,\,3$, and $1 \Mpc$, respectively.}
\label{fig:reclf}
\end{figure}
 
Obviously, transmission relies on the existence of large \hii regions.
By imposing a maximum effective bubble size $R_{\rm max}$,
recombinations eliminate the largest bubbles \citep{furl05-rec}.  We
can approximately incorporate this by assuming that the damping wing
absorption begins at $R_{\rm max}$ for any galaxy nominally in a
bubble larger than this radius (Fig.~9 of \citealt{furl05-rec} shows
this to be a reasonable approximation).  Figure~\ref{fig:reclf}
illustrates the effect on the luminosity function.  We show the
fraction of galaxies above a given luminosity threshold that are
visible relative to the case with $R_{\rm max}=\infty$.  We show a
range of values for $R_{\rm max}$, which is quite uncertain because of
our poor knowledge of the density structure of the IGM during
reionization.

A number of trends are apparent in Figure~\ref{fig:reclf}.  First,
recombinations only significantly affect $n(>L)$ if $R_{\rm max} \la 5
\Mpc$; $\tau_d$ is sufficiently small in larger bubbles that
restricting their sizes makes little difference.  Thus the luminosity
function will be unaffected if the density model of \citet{miralda00}
-- which essentially extrapolates the population of Lyman-limit
systems at $z \sim 3$ -- is accurate.  Second, $R_{\rm max}$ affects
brighter galaxies more strongly, because their mass function is
steeper and they reside in larger bubbles.  Third, $n(>L)$ is more
sensitive to $R_{\rm max}$ at $\bxio=0.75$ because more of the
universe is contained in bubbles with nominal $R>R_{\rm max}$.  That
trend reverses itself if $R_{\rm max} \ll R_c$, in which case the mean
neutral fraction in the absorbing gas dominates the effect (as in the
dotted curves).  (Note that this simple treatment may overestimate the
effect of imposing an $R_{\rm max}$: small recombination-limited
bubbles cluster strongly, rendering the absorption somewhat weaker
than our simple $\tau_d$ calculation would predict.)

\section{The Observed Correlation Function}
\label{bias}

FHZ04 pointed out that, in addition to decreasing the mean number
density of observable galaxies, the \hii regions will modify their
spatial distribution.  Here we show how this affects their clustering.
Dwarf galaxies could appear in a survey at $\bxio \ll 1$, but they
will be confined to the small volume filled by (rare) large bubbles.
On the other hand, each of these bubbles contains many galaxies. A
simple model illustrates how this enhances the apparent clustering on
small scales (relative to galaxies observed in the continuum, for
example).  Suppose that galaxies with number density $\bar{n}$ are
distributed randomly throughout the universe but that we can only
observe those with at least one neighbour within a sphere of
volume $V \ll \bar{n}^{-1}$.  Assuming a Poisson distribution, the
number density of observed objects would be $n_{\rm obs}=\bar{n} ( 1 -
e^{-\bar{n}V})$.  As usual the correlation function of the observed
sample is defined through the total probability of finding two
galaxies in volumes $\delta V_1$ and $\delta V_2$, $\delta P = n_{\rm
obs}^2 \, (1 + \xi) \, \delta V_1 \, \delta V_2$.  However, we know
that every observed galaxy has a neighbour within $V$; thus $\delta
P=n_{\rm obs} \, \delta V_1 \, (\delta V_2/V)$ for small separations
(where the factor $\delta V_2/V$ assumes the
neighbour to be randomly located within $V$).  Thus, $\xi = 1/(n_{\rm
obs} V) -1$ on such scales: even though the underlying distribution is
random, the selection criterion induces clustering.  Note that it can
be extremely large if $V \ll n_{\rm obs}^{-1}$.

On large scales, the modulation takes a different form.  An observed
galaxy resides in a large bubble, corresponding to an overdense
region.  
Because such regions are biased (i.e. more strongly clustered than the
underlying mass distribution), they will tend to lie near to
other overdense regions -- and hence to other large bubbles.  Thus, we
will be more likely to see galaxies near the original object than in
an average slice of the universe.  Because we do not see similar
galaxies in small (less-biased) bubbles, the large-scale bias will
generically be larger than that intrinsic to the galaxies.

Because these two effects have different amplitudes, the bubbles
introduce a scale-dependent bias to the correlation function of
galaxies, with a break at $r \approx R_c$.  Unfortunately, developing
a self-consistent analytic model for the correlation function $\xi(r)$
is difficult because spherical bubbles introduce spurious features on
scales $\sim R_c$, just the regime in which the scale dependence is
most interesting.  These features occur because of the difficulty of
efficiently filling space with non-overlapping spheres (see the
discussion in \citealt{mcquinn05}).  Instead we will compute the
limiting cases of $r \ll R_c$ and $r \gg R_c$.

By analogy with the halo model for the density field \citep{cooray02},
these limiting regimes correspond to correlations between galaxies
within a single bubble and within two separate bubbles.  We begin with
large scales: the observed clustering is the average bias of the
bubbles weighted by the number of galaxies in each \hii region
(analogous to the two-halo term for the density field):
\bqa
b_{r=\infty} & = & \int \deriv m_b \, n_b(m_b) \, b_b(m_b) \, V_b \nonumber \\ 
& & \times \int_{m_{h,{\rm min}}}^{m_b/\zeta} \deriv m_h \, \frac{n_h(m_h|m_b)}{\bar{n}_{\rm gal}},
\label{eq:blarge}
\eqa
where we integrate only over those haloes visible after damping wing
absorption and $\bar{n}_{\rm gal}$ is the mean number density of
observable galaxies.  Following the procedure of \citet{mo96}, the
bias $b_b$ of \hii regions is \citep{mcquinn05}
\bq
b_b(m_b) = 1 + \frac{B(m_b)/\sigma^2(m_b) - 1/B_0}{D(z)},
\label{eq:biasb}
\eq
where $D(z)$ is the linear growth factor.  
Here and throughout we work with the linear bias $b$, which is defined
by $n(m|\delta) = n(m)[1 + b \delta + {\mathcal O}(\delta^2)]$ for a
population with mean number density $n(m)$.  Note that (unlike the
halo bias) we can have $b_b < 0$: late in reionization, small bubbles
are truly \emph{anti}-biased because dense regions have already been
incorporated into large ionized regions.  Unfortunately, as discussed
by \citet{mcquinn05}, our linear bias formula breaks down late in
reionization, essentially because the physical requirement that $x_i
\leq 1$ caps the effective number density and hence the bias.  By
comparing to the FZH04 ionization model as implemented in numerical
realisations of gaussian random fields, \citet{mcquinn05} found that
equation (\ref{eq:biasb}) overpredicts the biasing when $\bxio \sim
1$.  Our quantitative predictions for large $\bxio$ should therefore
be taken with caution.

The behaviour on small scales is somewhat more subtle.  If galaxies
were randomly distributed within each bubble, the simple argument in
the first paragraph of this section suggests that the correlation
function would just be the weighted average of the number of pairs
per \hii region.  However, in addition to the increase in the number
of galaxies in each bubble, the galaxies also trace density
fluctuations within each bubble.  We therefore write
\bq
b_{\rm sm}^2 = \int \deriv m_b \, n_b(m_b) \, V_b \, b_h^2(m_b) \, \frac{\langle N_{\rm gal}(N_{\rm gal}-1) | m_b \rangle}{\bar{N}_{\rm gal}^2},
\label{eq:bsmall}
\eq
where $\bar{N}_{\rm gal}= \bar{n}_{\rm gal} V_b$, $\langle N_{\rm
gal}(N_{\rm gal}-1) | m_b \rangle$ is the expected number of galaxy
pairs within each bubble, and $b_h^2$ measures the excess bias of
these haloes inside each bubble.\footnote{This expression can be
derived formally (modulo the precise definition of $b_h^2$; see below)
by constructing the galaxy density field from bubbles and their
constituent haloes, in analogy to the halo model.  It corresponds to
the ``two-halo, one-bubble" term in such a treatment; i.e.,
correlations between two particles that lie in the same bubble but
different dark matter haloes.  The ``bubble profile" describing the
distribution of galaxies within the bubble turns out to be
proportional to the square root of the linear matter correlation
function.}  We note that equation (\ref{eq:bsmall}) ignores nonlinear
corrections to the density field (the one-halo term), which dominate
on scales of tens of kpc at these redshifts.  It therefore applies
only to separations intermediate between this nonlinear scale and
$R_c$.

We can write the expected number of pairs as
\bqa
\langle N_{\rm gal}(N_{\rm gal}-1) | m_b \rangle & \approx & V_b \int \deriv m_h \, n_h(m_h|m_b) \, (V_b-\zeta m_h/\bar{\rho}) \nonumber \\ 
& & \times \int^{m_{\rm max}} \deriv m_h' \, n_h(m_h'|m_b),
\label{eq:galpair}
\eqa
where the factor in parentheses accounts for the ``bubble exclusion
effect": the total mass of galaxies within a specified bubble cannot
exceed $m_b/\zeta$ or we would overproduce the number of
ionizing photons.  This also sets $m_{\rm max}=m_b-\zeta m_h$.
Because most galaxies have $m_h \sim \mmin \ll m_b/\zeta$, we will
take $m_{\rm max}=m_b/\zeta$ but begin the integration over $m_b$ in
equation (\ref{eq:bsmall}) at
$2 \zeta \mmin$ in order to exclude bubbles that can contain only a
single galaxy.  With these simplifications,
\bq
\langle N_{\rm gal}(N_{\rm gal}-1) | m_b \rangle \approx {\rm max}\{0,\bar{N}_{\rm gal}(m_b) [ \bar{N}_{\rm gal}(m_b) -1]\}.
\label{eq:galpairsimple}
\eq
This approximation is reasonable so long as most of the observed
galaxies are in bubbles with $\bar{N}_{\rm gal} \ga 2$; if this is not
true, then we would underestimate the clustering because the pair
count is dominated by small bubbles with an unusual galaxy
distribution.  This only happens if the luminosity threshold in the
survey is extremely large or at quite high redshifts (for example, a
survey at $z=15$ with $m_{\rm obs,min}=10^{10} \msun$ encounters these
problems).
In particular, our approximation is valid for all the cases we show here.
Of course, the first high-redshift surveys may only see the brightest
galaxies and so lie within the regime where our formalism breaks
down.  Properly interpreting small-scale clustering in such cases
requires more sophisticated methods, in particular numerical
simulations or hybrid approaches similar to \citet{zahn05}.

The remaining factor is $b_h(m_b)$.  It may seem reasonable to take
this to be the mean value of the usual \citet{mo96} halo bias,
evaluated over $n_h(m_h|m_b)$.  However, the pair density inside each
bubble \emph{already} includes much of this bias because it counts the
number of galaxies in a region with overdensity $\delta_x$.  We
therefore only want the ``excess" bias of the galaxies relative to
density fluctuations on scales smaller than $m_b$, which is the bias
evaluated from the conditional mass function in equation
(\ref{eq:nhcond}).  Following the procedure of \citet{mo96}, we find
\bq
b_h(m_h|m_b) = 1 + \frac{(\delta_c-\delta_x)^2/(\sigma^2-\sigma_b^2) - 1}{\delta_c(z=0) - \delta_x(z=0)}.
\label{eq:bcond}
\eq
Because the presence of the first galaxy in a pair biases the second
to be smaller than average (the bubble exclusion effect mentioned
above), we take
\bq
b_h^2(m_b) \approx \langle b_h|m_b \rangle \, b_h(m_{h,{\rm min}}),
\label{eq:bh2}
\eq
where $\langle b_h|m_b \rangle$ is the average of equation
(\ref{eq:bcond}) over $n_h(m_h|m_b)$.  This will slightly
underestimate the true bias by forcing all of the galaxy's neighbours
to be of the minimum possible size (and hence bias).

\begin{figure}
\begin{center}
\resizebox{8cm}{!}{\includegraphics{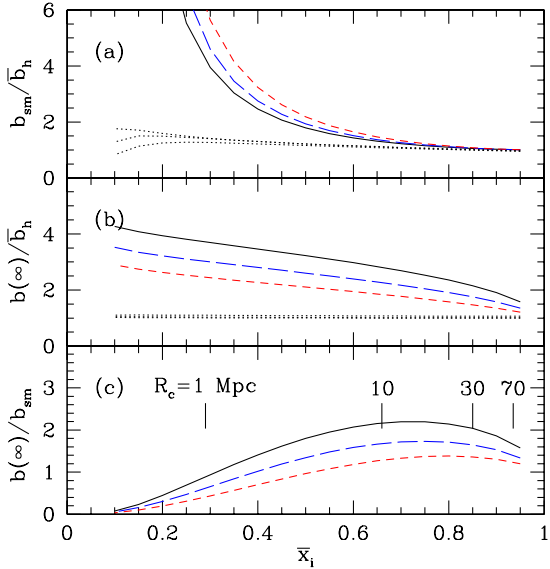}}\\%
\end{center}
\caption{\emph{(a)}: Predicted small-scale bias at $z=10$, relative to
the bias expected if all galaxies above the mass threshold were
visible.  This applies to separations larger than the nonlinear scale
but smaller than $R_c$.  The solid, long-dashed, and short-dashed
curves take $m_{\rm obs,min}=m_4, \, 10^9$, and $10^{10} \msun$,
respectively.  The dotted curves show the predicted
\emph{absorption-free} galaxy bias relative to its true value (see
text).  \emph{(b)}: Predicted large-scale bias at $z=10$, relative to
the bias expected if all galaxies above the mass threshold were
visible.  \emph{(c)}: Ratio of large to small scale bias.  We mark the
characteristic bubble size $R_c$ (in Mpc) as well.}
\label{fig:z10bias}
\end{figure}

We show the resulting bias at $z=10$ (as a function of $\bxio$) in
Figure~\ref{fig:z10bias}.  In each panel, the solid, long-dashed, and
short-dashed curves take $m_{\rm obs,min}=m_4,\, 10^9 \msun$, and
$10^{10} \msun$, respectively.  Panels \emph{(a)} and \emph{(b)} show
$b_{\rm sm}$ and $b_{r=\infty}$.  We scale the results to the bias
$\bar{b}_h$ intrinsic to the galaxy population if absorption could be
ignored.  Panel \emph{(c)} shows the ratio $b_{r=\infty}/b_{\rm sm}$,
illustrating the magnitude of the ``break" in the linear bias.  We
emphasize that the scale at which the break occurs will evolve
throughout reionization along with the characteristic bubble size
$R_c$; for illustrative purposes we mark several values of $R_c$.

Before commenting on our results, we note that the approximations in
the FZH04 model, equation (\ref{eq:biasb}), and the treatment of
galaxy pairs will obviously cause errors in the predicted bias.
Fortunately, we can check these by noting that we should recover
$\bar{b}_h = b_{\rm sm} = b_{r=\infty}$ if we neglect absorption
entirely by including \emph{all} galaxies in each bubble.  Here
$\bar{b}_h=\int \deriv m \, n_h(m) \, b_h(m)/\int \deriv m \, n_h(m)$
is the actual mean bias of the galaxy population.  The dotted curves
in panels \emph{(a)} and \emph{(b)} compare the recovered values to
the true bias.  Without absorption, equation (\ref{eq:blarge}) is
equivalent to an integral over the halo mass function; the bubble bias
is essentially the same as the average bias of the haloes it contains,
so the resulting errors are typically $\la 10\%$ (and furthermore do
not vary much with $\bxio$).  The small-scale result in equation
(\ref{eq:bsmall}) becomes the weighted average of the bias of pairs of
galaxies in each bubble, where the galaxy bias is built from the
conditional $b_h(m_h|m_b)$ and the number of pairs.  Panel \emph{(a)}
shows that the formalism is accurate for moderate to large $\bxio$ but
encounters problems when $\bxio$ is small.  At worst this error is a
factor $\sim 2$.  We are primarily interested in the observable bias
relative to the intrinsic value, so we attempt to remove this error by
scaling all of our curves in units of the predicted value, rather than
the true intrinsic bias.  Note, however, that it makes a significant
difference only for $\bxio$ near the minimum value shown in each plot.

Clearly, both $b_{\rm sm}$ and $b_{r=\infty}$ decrease throughout
reionization.  The large-scale bias decreases because the ionized
regions must lie nearer to the mean density (and hence be less biased)
as $\bxio \rightarrow 1$: this behaviour must be generic to any model
in which reionization begins in overdense regions.  The small-scale
bias decreases because bubbles large enough to allow transmission
become common: early on, only those galaxies with near neighbours are
visible, so the correlations are strong.  In the middle and final
stages of reionization, most galaxies lie inside bubbles large enough
to permit transmission, so more typical galaxies become visible and
$b_{\rm sm} \rightarrow \bar{b}_h$.

Note that $b_{r=\infty}/\bar{b}_h$ decreases with $m_{\rm obs,min}$ in
panel \emph{(b)}.  This may seem counterintuitive, because more massive
galaxies are also more highly biased.  In fact it is true that the
value of $b_{r=\infty}$ increases with $m_{\rm obs,min}$, but the
amplification relative to the underlying population is larger for
small galaxies.  Large galaxies preferentially sit in large bubbles
and so suffer relatively little absorption.  But a large fraction of
small galaxies reside in small bubbles, where they disappear because
of damping wing absorption.  Thus the \lya selection picks out a more
unusual population of small galaxies than it does for large galaxies,
explaining the former's increased amplification.
If the resonant absorption decreases with halo mass (see the
discussion in \S \ref{lumfcn}), that would exaggerate this trend,
because faint galaxies would have to sit in even larger bubbles.  (For
the same reason, it would also increase the small-scale bias.)

As emphasized above, the transition between these two regimes occurs
on scales $r \approx R_c$.  Measuring the location of the break will
thus directly constrain $n_b(m)$.  Interestingly, our model also shows
significant evolution in $b_{r=\infty}/b_{\rm sm}$ throughout
reionization, offering an independent measure of $\bxio$.  Early in
reionization, the small-scale bias dominates because large bubbles are
extremely rare (a survey would find a few distinct clumps of galaxies,
even though the underlying galaxy distribution is fairly uniform; the
clustering on scales smaller than the average clump thus appears
enormous).  However, once a typical bubble grows enough to allow
significant transmission, the small-scale bias rapidly falls to that
intrinsic to the haloes (compare to Figure~\ref{fig:td}\emph{b}).
Meanwhile, the large-scale bias remains substantial (subject to the
caveat that our linear treatment eventually breaks down) -- these
bubbles are overdensities on scales much larger than the typical
galaxy, so they cluster more strongly.  Thus toward the end of
reionization $b_{r=\infty}$ is typically larger by a factor of order
unity; the ratio increases slowly as $z$ increases.

\begin{figure}
\begin{center}
\resizebox{8cm}{!}{\includegraphics{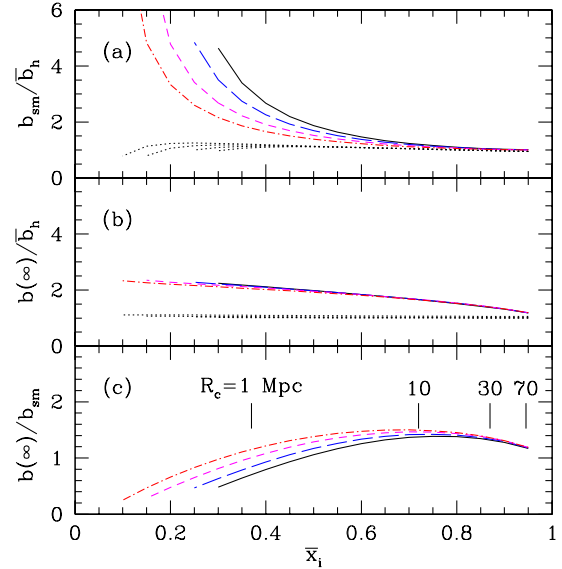}}\\%
\end{center}
\caption{As Fig.~\ref{fig:z10bias}, but for $z=6.5$.  All the curves
take $m_{\rm obs,min}=10^{10} \msun$.  The solid, long-dashed,
short-dashed, and dash-dotted curves take $\mmin/m_4=1,\,3,\,10$, and
$30$, respectively.  We mark $R_c$ for $\mmin=m_4$ here; see
Fig.~\ref{fig:td}\emph{a} for $R_c$ in the other scenarios. }
\label{fig:z6bias}
\end{figure}

Figure~\ref{fig:z6bias} shows the bias for a survey with $m_{\rm
obs,min}=10^{10} \msun$ at $z=6.5$.  Here, we vary the minimum halo
mass to host a galaxy, $\mmin$.  Comparing the solid curves in
Figures~\ref{fig:z10bias} and \ref{fig:z6bias}, we find that the
induced large-scale bias decreases with cosmic time, but the small-scale
bias does not change dramatically.  Interestingly, the mass threshold
has virtually no effect on the large-scale bias: although $n_b(m)$
shifts to larger scales, the bias averaged over each distribution
remains nearly constant.  However, by increasing $R_c$, the modulation
becomes less severe and $b_{\rm sm}$ decreases at a fixed $\bxio$.
Thus the small-scale bias early in reionization is also sensitive to
the bubble size distribution.

\section{Discussion}
\label{disc}

We have shown that the sizes of \hii regions during reionization
strongly affect the distribution of \lyans-selected galaxies (see also
FHZ04).  The neutral IGM extinguishes the \lya lines of galaxies,
reducing their number counts.  Without source clustering, their
abundance declines significantly when $\bxio \la 0.7$--$0.9$, with the
range depending on assumptions about the detailed intrinsic properties
of the galaxies \citep{santos04}.  In our model, the existence of
large \hii regions implies that their abundance only begins to decline
when $\bxio \la 0.5$.  This weakens existing constraints on the
neutral fraction at $z=6.5$ \citep{malhotra04}, but it also implies
that the \lya selection technique will be useful even into the middle
stages of reionization.

Moreover, we have shown that the extinction is not uniform across the
sky -- galaxies in large \hii regions remain visible, but those in
smaller bubbles vanish.  The bubbles thus induce extra clustering in
the galaxy distribution.  On large scales, the observed galaxies trace
the highly-biased large bubbles.  On small scales, the clustering
increases rapidly at early times because only those exceptional
regions with an excess of galaxies are visible.  As a result, we
expect a break in the observed bias to appear on the typical scale of
the \hii regions, with a magnitude that evolves throughout
reionization.  Thus large area surveys for \lyans-emitters have the
potential to offer powerful constraints on the topology of neutral gas
during reionization.

One potential obstacle to interpreting the luminosity function and
clustering measurements is our poor knowledge of the intrinsic
properties of these galaxies.  The luminosity function test is
clearest as a differential measurement between two nearby redshifts
(as indeed \citealt{malhotra04} applied it), because that decreases
the amount of evolution one would expect in the intrinsic source
population.  However, if reionization is extended, such differential
tests may not offer much insight.  Clustering can encounter similar
difficulties: lower redshift \lya galaxies are highly clustered
\citep{hamana04,ouchi05}, and it may seem difficult to extract
information about reionization without understanding their intrinsic
properties.  In particular, the density of \lyans-selected galaxies at
$z=5.7$ shows rather large variations on large scales
\citep{ouchi05}. More effort is clearly needed to understand these
systems, and their lower-redshift analogues, in greater depth.

Fortunately, there is one strategy that may allow us to sidestep these
difficulties: we can compare \lyans-selected samples to samples
collected through other techniques, such as broadband Lyman-break
selection.  Provided that the latter can be corrected for the effects
of the \lya line properties on continuum selection (which may not be
trivial) and that the average line properties of galaxies of a given
luminosity do not evolve rapidly with redshift, they can be used as a
control sample.  For example, if the number counts of \lyans-selected
galaxies decline several times faster than Lyman-break galaxies over
some redshift interval, that would constitute strong evidence for a
significant change in $\bxio$.  The Lyman-break sample can also be
used to estimate the evolution of the intrinsic halo bias as a control
for measurements of the small and large-scale bias induced by the
bubbles.  (Note that the clustering also offers another consistency
check: the intrinsic galaxy bias should be independent of separation,
but \lya absorption induces a scale-dependent feature that evolves
throughout reionization.)  Thus, the \lya and Lyman-break selection
techniques nicely complement each other.  There are three caveats to
this approach, however,  First, the \lya line may have a non-trivial
effect on the photometric selection.  Second, there is some weak
evidence that \lya line properties do evolve at lower redshifts,
although this would only be a problem if the evolution occurs on a
shorter timescale than changes in $\bxio$.  Finally,
continuum-selection techniques often have difficulty reaching similar
depths to line-selection, which will restrict the available range of
line luminosities.

Our models are extremely simple, and one can imagine any number of
complications.  However, our qualitative conclusions are robust.  For
example, we have assumed the ionizing efficiency of galaxies to be
independent of their mass.  \citet{furl05-charsize} show how to relax
this assumption within the context of the FZH04 model.  They find that
making $\zeta$ an increasing function of halo mass will increase $R_c$
because it boosts the average bias of the underlying galaxy
population.  Thus the damping wing absorption decreases at a given
$\bxio$, allowing \lya selection to penetrate even deeper into the
reionization epoch.  For instance, if $\zeta \propto m^{1/3}$ at
$z=6.5$, half the galaxies disappear when $\bxio \sim 0.5$ (compared
to half at $\bxio \sim 0.75$ if $\zeta=$constant; see
Fig.~\ref{fig:z6lf}).  Such an assumption also tends to reduce the
boost in small-scale bias at early times, but it does not affect the
existence of the break in clustering at $\sim R_c$.

We also showed that incorporating recombinations can significantly
reduce the number counts of \lyans-selected galaxies by imposing a
maximum size on the bubbles, but only if the IGM is much clumpier than
an extrapolation from $z\sim 3$ would imply \citep{miralda00}.  Such
clumpiness may not be surprising, because Jeans smoothing is much less
effective if the gas remains cold until photoionization (e.g.,
\citealt{iliev05}).  It will also affect the clustering: in such a
picture large bubbles require a substantial excess of sources,
boosting the small-scale bias by an even larger amount than in the
FZH04 model.  Furthermore, it affects the characteristic size of the
\hii regions and so changes the location of the expected break in the
clustering strength.

Our analytic models are only approximate, and our predictions can be
improved with numerical techniques, even short of full cosmological
simulations.  For example, applying the FZH04 ionization criterion to
numerical simulations or gaussian random fields (as in
\citealt{zahn05}) will allow us to include the distribution of
galaxies within each bubble, the (non-spherical) morphology of
bubbles, their proper biasing for large $\bxio$, and analyze galaxy
samples restricted to the brightest objects.  Such an approach will
yield predictions for the full, scale-dependent correlation function
of \lyans-selected galaxies, which will allow much more detailed
comparisons with future observations.  Complete numerical simulations
with radiative transfer will allow even more sophisticated tests of
the model, including our treatment of recombinations, source
clustering, and the intrinsic clumpiness of the IGM.

Our results have a number of implications for ongoing and future
searches for high-redshift \lyans-emitters.  First and foremost, our
robust prediction of large \hii regions implies that this selection
technique will allow us to probe farther back in the reionization
process than one might have naively expected.  Thus, it is crucial to
push these surveys to the highest redshifts possible.  Once galaxies
are detected at a particular redshift, we advocate extending the
search to as wide a contiguous area as possible in order to measure
the clustering.  Even though constructing a catalogue sufficiently
deep to measure the large-scale bias may be impractical given
telescope time constraints, any detection of extremely large
small-scale clustering would offer strong evidence for a relatively
large neutral fraction.  Studies like these, combined with other
probes of the high redshift IGM such as 21 cm emission (e.g.
Zaldarriaga et al. 2004), therefore have the potential to reveal
how reionization developed and to elucidate the sources responsible for
this process (e.g. Furlanetto et al. 2004c).

\vspace{0.2in}

We thank an anonymous referee for helpful comments that improved the
manuscript.  This work was supported by NSF grants ACI 96-19019, AST
00-71019, AST 02-06299, and AST 03-07690, and NASA ATP grants
NAG5-12140, NAG5-13292, and NAG5-13381.


\end{document}